# EXCITATION OF THE TRANSITION RADIATION BY A RELATIVISTIC ELECTRON BUNCH IN THE PLASMA HALF-SPACE

***Balakirev V.A., Onishchenko I.N.***
***KIPT, Kharkov, Ukraine, E-mail:onish@kipt.kharkov.ua***

The process of transition excitation of an electromagnetic field by a relativistic electron bunch in a plasma half-space is investigated. The influence of the plasma boundary on the spatial structure of plasma wake oscillations is studied. The shape and intensity of the transition electromagnetic pulse propagating in the plasma are determined.

## INTRODUCTION

The processes of excitation of wake Langmuir oscillations in plasma by relativistic electron bunches, as a rule, were considered on the models of plasma systems unbounded in the longitudinal direction [1-9]. Accounting of the transversal (radial) boundaries of plasma, as an example for plasma waveguides of various geometries, was carried out in [9-11]. Meanwhile, the presence of the longitudinal boundaries of plasma can lead to a number of qualitative features of the excitation picture of plasma wake fields by both relativistic electron bunches or laser pulses.

In the present work, within the framework of the plasma half-space model, the influence of a sharp plasma boundary on the excitation picture of wake plasma oscillations by a relativistic electron bunch is studied.

In the system under consideration beside the longitudinal Langmuir oscillations, a relativistic electron bunch will excite too the transition electromagnetic pulse [12-17]. In work attention is also paid to the spatial-temporal structure of the transition electromagnetic pulse and its intensity and energy.

Note that the intense transition electromagnetic pulse of ultrashort duration can serve, in particular, as a tool for diagnostics of the parameters of plasma and electron bunches.

## 1. STATEMENT OF THE PROBLEM. BASIC EQUATIONS

Plasma half-space $z>0$ ($z$ is longitudinal coordinate) is bounded by a perfectly conducting plane $z=0$. An axisymmetric relativistic electron bunches is injected into the plasma from a perfectly conducting plane normal to it. Our task is the determination of the electromagnetic field, including the wake Langmuir oscillations, excited in a semi-bounded plasma by an relativistic electron bunch.

The initial system of equations contains Maxwell's equations

$$rot\vec{E} = -\frac{1}{c}\frac{\partial \vec{H}}{\partial t}, \quad rot\vec{H} = \frac{1}{c}\frac{\partial \vec{D}}{\partial t} + \frac{4\pi}{c}\vec{j}_b, \quad (1)$$

$$div\vec{D} = 4\pi\rho, \quad div\vec{H} = 0,$$

$\rho_b, \vec{j}_b$ are charge density and current of an electron bunch, $\vec{D} = \hat{\varepsilon}\vec{E}$ is electric displacement field, $\hat{\varepsilon}$ is dielectric constant operator of a plasma.

Maxwell's system of equations (1) describes the excitation of an electromagnetic field by external charges and currents in a plasma. On the surface of a perfectly conducting plane the tangential component of the electric field vanishes

$$\vec{E}_t = 0. \quad (2)$$

We will solve the problem of wake field excitation by an axisymmetric electron bunch in a plasma as follows. First, we determine the electromagnetic field $\vec{E}_G, \vec{H}_G$ (Green's function) of a moving charge in the form of an infinitely thin ring with a charge density

$$d\rho = -dQ\frac{1}{v_0}\frac{\delta(r-r_0)}{2\pi r}\delta(t-\frac{z}{v_0}-t_0), \quad 3)$$

where $r$ is radial coordinate, $r_0$ is ring radius, $t_0$ is time of entry of an elementary ring bunch into the plasma, $v_0$ is bunch velocity, $dQ(r_0,t_0)$ is elementary charge of the ring connected with the current density of the bunch at the entrance to the plasma $(z=0)$ $j_0(t_0,r_0)$ by the relation

$$dQ = j_0(t_0,r_0)2\pi r_0 dr_0 dt_0.$$

The current density of an elementary ring charge is determined by the expression

$$d\vec{j} = v_0 d\rho \vec{e}_z, \quad (4)$$

$\vec{e}_z$ is unit vector in longitudinal direction.

Let's consider an electron bunch with the current density

$$j_0(r_0,t_0) = j_0 R(r_0/r_b) T(t_0/t_b), \quad (5)$$

where the function $R(r_0/r_b)$ describes dependence of the bunch density on radius (transverse profile), $r_b$ is characteristic transverse bunch size, function $T(t_0/t_b)$ describes the longitudinal density profile of a bunch, $t_b$ is characteristic duration of the bunch. The value $j_0$ is connected with the full charge $Q$ by the relation $j_0 = Q/(s_{eff}t_{eff})$, where $s_{eff}$ is effective cross section of the bunch

$$s_{eff} = \pi r_b^2\hat{\sigma}, \ \hat{\sigma} = 2\int_0^\infty R(\rho_0)\rho_0 d\rho_0,$$

and $t_{eff}$ is effective bunch duration

$$t_{eff} = \hat{\tau}t_b. \ \hat{\tau} = \int_{-\infty}^{\infty} T(\tau_0)d\tau_0.$$

If we define the electromagnetic field excited by an elementary ring charge (3) and current (4)

$$\vec{E}_G(r,r_0,z,t-t_0) = dQ\vec{\mathrm{E}}(r,r_0,z,t-t_0), \quad (6)$$

where $\vec{\mathrm{E}}(r, r_0, z, t-t_0)$ is the electric field excited by a ring electron bunch with a unit charge, then the total electromagnetic field excited by an electron bunch of finite sizes may be found by summing (integrating) the fields of elementary ring bunches

$$\vec{E}(r,z,t) = \int_0^b 2\pi r_0 dr_0 \int_{-\infty}^{t} dt_0\, j(r_0,t_0)\vec{\mathrm{E}}(r,r_0,z,t-t_0)\,.$$

Taking into account relations (5), (6), this expression can be written as follows

$$\vec{E}(r,z,t) = \frac{2\pi Q}{s_{eff} t_{eff}} \int_0^b R\left(\frac{r_0}{r_b}\right) r_0 dr_0 \times$$

$$\times \int_{-\infty}^{t} T\left(\frac{t_0}{t_b}\right) \vec{\mathrm{E}}(r,r_0,z,t-t_0) dt_0\,. \qquad (7)$$

The next step of the solution problem is to determine the electromagnetic field (Green's function) (6) of an elementary ring electron bunch.

## 2. DETERMINATION OF THE GREEN FUNCTION

From the initial system of equations (1) follows the wave equation for the electric field excited by a thin ring electron bunch with charge (3) and current (4) densities (Green's functions $\vec{E}_G$ )

$$\Delta\vec{E}_G - \frac{1}{c^2}\frac{\partial^2}{\partial t^2}\hat{\varepsilon}\vec{E}_G = -\frac{2dQ}{v_0}\left[\frac{v_0}{c^2}\vec{e}_z \frac{\delta(r-r_0)}{r}\frac{\partial}{\partial t}\delta\left(t-t_L\right) + \right.$$

$$\left. + \hat{\varepsilon}^{-1}\vec{\nabla}\left(\frac{\delta(r-r_0)}{r}\delta\left(t-t_L\right)\right)\right],\ t_L = t_0 + \frac{z}{v_0}\,, \qquad (8)$$

$\hat{\varepsilon}^{-1}$ is operator inverse to $\hat{\varepsilon}$. Due to the symmetry of the system, the electron bunch will excite electromagnetic waves with field components $E_z, E_r, H_\varphi$. From the vector inhomogeneous wave equation (8) follows the equation for the radial component of the electric field

$$\frac{1}{r}\frac{\partial}{\partial r} r \frac{\partial}{\partial r} E_{Gr} - \frac{1}{r^2}E_{Gr} + \frac{\partial^2}{\partial z^2}E_{Gr} - \frac{1}{c^2}\frac{\partial^2}{\partial t^2}\hat{\varepsilon}E_{Gr} = .$$

$$= -\hat{\varepsilon}^{-1}\frac{2dQ}{v_0}\delta\left(t-t_L\right)\frac{\partial}{\partial r}\left(\frac{\delta(r-r_0)}{r}\right). \qquad (9)$$

Equation (9) must be supplemented with the boundary condition

$$E_{Gr}(z=0) = 0\,.$$

We will solve the wave equation (9) by the Fourier-Bessel transform method

$$E_{Gr}(z,r,t) = \int_0^\infty J_1(\lambda r)\lambda d\lambda \int_{-\infty}^{\infty} \hat{E}_{Gr\omega\lambda}(z) e^{-i\omega t} d\omega, \quad (10)$$

$$\hat{E}_{Gr\omega\lambda}(z) = \frac{1}{2\pi}\int_0^\infty J_1(\lambda r)\lambda d\lambda \int_{-\infty}^{\infty} E_{Gr}(z,r,t) e^{i\omega t} dt \qquad .$$

From the wave equation (9) the ordinary differential equation for the Fourier-Bessel component $\hat{E}_{Gr\omega\lambda}(z)$ of the radial electric field follows

$$\frac{d^2}{dz^2}\hat{E}_{Gr\omega\lambda} + k_z^2 \hat{E}_{Gr\omega\lambda} = \frac{dQ}{\pi v_0 \varepsilon(\omega)} e^{i(k_l z + \omega t_0)k}\lambda J_0(\lambda r_0)\,,$$

where

$$k_l = \omega / v_0,\quad k_z(\omega) = \sqrt{k_0^2\varepsilon(\omega) - \lambda^2}\,,\ k_0 = \omega / \quad ,$$

$$\varepsilon(\omega) = 1 - \frac{\omega_p^2}{\omega(\omega + i\nu)}$$

is the dielectric constant of the plasma, $\omega_p$ is the Langmuir frequency of the plasma, and $\nu$ is the effective frequency of collisions. The solution of this equation, which satisfies the boundary condition on an ideally conducting plane $\hat{E}_{Gr\omega\lambda}(z=0) = 0$ and the condition of radiation at infinity $(z \to \infty)$, has the form

$$\hat{E}_{Gr\omega\lambda} = dQ\frac{\lambda J_0(\lambda r_0) e^{i\omega t_0}}{\pi v_0 \varepsilon(\omega)\left(k_z^2 - k_l^2\right)}\left[e^{ik_l z} - e^{ik_z z}\right].$$

Taking into account the Fourier-Bessel integral representation (10), the expression for the radial component of the electric field can be written as follows

$$E_{Gr} = \frac{dQ}{\pi v_0}\int_0^\infty \Pi_1(\lambda r)\lambda^2 d\lambda \int_{-\infty}^{\infty} \frac{e^{-i\omega(t-t_L)}}{\varepsilon(\omega)\left(k_z^2 - k_l^2\right)} d\omega -$$

$$-\frac{dQ}{\pi v_0}\int_0^\infty \Pi_1(\lambda r)\lambda^2 d\lambda \int_{-\infty}^{\infty} \frac{e^{-i[\omega\bar{t} - k_z(\omega)z]}}{\varepsilon(\omega)\left(k_z^2 - k_l^2\right)} d\omega\,, \qquad (11)$$

$$\Pi_n(\lambda r) = J_n(\lambda r)J_0(\lambda r_0),\ n = 0;1,\ \ \bar{t} = t - t_0\,.$$

From Maxwell's equations (1), the expressions follow for the other components of the electromagnetic field

$$E_{Gz} = i\frac{dQ}{\pi c}\int_0^\infty \Pi_0(\lambda r)\lambda d\lambda \int_{-\infty}^{\infty} \frac{e^{-i\omega(t-t_L)}}{k_0\varepsilon(\omega)}\frac{k_0^2\varepsilon - k_l^2}{k_z^2 - k_l^2} d\omega -$$

$$-i\frac{dQ}{\pi c}\int_0^\infty \Pi_0(\lambda r)\lambda^3 d\lambda \int_{-\infty}^{\infty} \frac{k_l e^{-i[\omega\bar{t} - k_z(\omega)z]}}{k_0 k_z \varepsilon(\omega)\left(k_z^2 - k_l^2\right)} d\omega, \qquad (12)$$

$$H_{G\varphi} = \frac{dQ}{\pi c}\int_0^\infty \Pi_1(\lambda r)\lambda^2 d\lambda \int_{-\infty}^{\infty} \frac{e^{-i\omega(t-t_L)}}{k_z^2 - k_l^2} d\omega -$$

$$-\frac{dQ}{\pi c}\int_0^\infty \Pi_1(\lambda r)\lambda^2 d\lambda \int_{-\infty}^{\infty} \frac{k_l e^{-i[\omega\bar{t} - k_z(\omega)z]}}{k_z\left(k_z^2 - k_l^2\right)} d\omega\,. \qquad (13)$$

The first terms in the expressions for the components of the electromagnetic field (11)-(13) describe the electromagnetic field $\vec{E}_{GB}, \vec{H}_{GB}$ of an electron ring bunch in unbounded plasma. The second terms in these formulas describe the field that arose due to the presence of the plasma boundary (i.e. transition electromagnetic field). Integrands of the electric fields components $\vec{E}_{GB}$ contain only simple poles

$$\omega = \pm\omega_p - i\nu / 2 \qquad (14)$$

located in the lower half-plane of the complex variable $\omega$ near the real axis, as well as the poles located on the imaginary axis

$$\omega = i\left(\pm\omega_{st} - \frac{1}{2}\frac{\omega_\nu^3}{\omega_{st}^2}\right), \qquad (15)$$

where

$$\omega_{st} = \beta_0\gamma_0\sqrt{\omega_p^2 + \lambda^2 c^2}\,,\ \omega_\nu = \left(\beta_0^2\gamma_0^2\omega_p^2\nu\right)^{1/3},$$

$$\beta_0 = v_0 / c,\ \gamma_0 = 1/\sqrt{1-\beta_0^2}\,.$$

Taking into account the weak dissipation of energy in the plasma gives only a rule for bypassing the singular points and we will not take it into account in the final formulas. After calculation of the residues in these poles, we find expressions for the electromagnetic field of a ring electron bunch in unbounded plasma

$$\vec{E}_{GB} = \vec{E}_{GB}^{(st)} - \nabla\Phi_{Gw}\,,\ H_{GB\varphi} = H_{GB\varphi}^{(st)}. \qquad (16)$$

Here

$$E_{GBz}^{(st)} = dQk_p^2\gamma_0^2 sign(\varsigma_L - \varsigma)\int_0^\infty \Pi_0(\lambda\eta)\frac{\lambda^3 e^{-\kappa|\varsigma-\varsigma_L|}}{\kappa^2+1}d\lambda, \quad (17)$$

$$E_{GBr}^{(st)} = -dQk_p^2\gamma_0^2\int_0^\infty \Pi_1(\lambda\eta)\frac{\lambda^2\kappa e^{-\kappa|\varsigma-\varsigma_L|}}{\kappa^2+1}d\lambda\,, \quad (18)$$

$$H_{GB\varphi}^{(st)} = -dQk_p^2\beta_0\gamma_0^2\int_0^\infty \Pi_1(\lambda\eta)\frac{\lambda^2 e^{-\kappa|\varsigma-\varsigma_L|}}{\kappa}d\lambda. \quad (19)$$

In these expressions we use the following dimensionless parameters and variables

$$\eta = k_p r,\ \eta_0 = k_p r_0,\ k_p = \omega_p / v_0,\ \varsigma = k_p z,\ \varsigma_L = k_p z_L,$$

$$z_L = v_0(t - t_0)\,,\ \kappa = \gamma_0\sqrt{\lambda^2+\beta_0^2}\,,\ \lambda \to \lambda / k_p\,.$$

Components (17)-(19) describe the quasi-static electromagnetic field of a ring electron bunch moving uniformly and rectilinearly in unbounded plasma. The second term in the expression for the electric field (16) corresponds to the purely potential wake field of Langmuir oscillations

$$\Phi_{Gw} = 2dQk_p\Gamma_0(\eta,\eta_0)\vartheta(\tau-\tau_L)\sin(\tau-\tau_L)\,, \quad (20)$$

$$\Gamma_0(\eta,\eta_0)\begin{cases} I_0(\eta_0)K_0(\eta), & \eta \ge \eta_0, \\ I_0(\eta)K_0(\eta_0), & \eta \le \eta_0, \end{cases}$$

$\vartheta(\tau-\tau_L)$ is Heaviside unit function,

$\tau = \omega_p t,\ \tau_L = \varsigma + \tau_0,\ \tau_0 = \omega_p t_0\,.$

Let us to study the transition electromagnetic field excited in the plasma half-space by a ring electron bunch. So, for example, for the transverse components of the transition electromagnetic field from expressions (11) - (13) we have

$$E_{GTr} = -\frac{dQ}{\pi v_0}\int_0^\infty \Pi_1(\lambda r)F_{Tr}(\bar{t},z,\lambda)\lambda^2 d\lambda\,, \quad (21)$$

$$F_{Tr}(\bar{t},z,\lambda) = \int_{-\infty}^{\infty} F_{Tr\omega}(z,\lambda)e^{-i[\omega\bar{t}-k_z(\omega)z]}d\omega, \quad (22)$$

$$F_{Tr\omega}(z,\lambda) = \frac{1}{\varepsilon(\omega)\left(k_z^2-k_l^2\right)},$$

$$H_{GT\varphi} = -\frac{dQ}{\pi c}\int_0^\infty \Pi_1(\lambda r)F_{T\varphi}(\bar{t},z,\lambda)\lambda^2 d\lambda\,, \quad (23)$$

$$F_{T\varphi}(\bar{t},z,\lambda) = \int_{-\infty}^{\infty} F_{T\varphi\omega}(z,\lambda)e^{-i[\omega\bar{t}-k_z(\omega)z]}d\omega\,, \quad (24)$$

$$F_{T\varphi\omega}(z,\lambda) = \frac{k_l e^{-i[\omega\bar{t}-k_z(\omega)z]}}{k_z\left(k_z^2-k_l^2\right)}\,.$$

Beside the simple poles (14), (15), the integrand in (22) has branch points

$$\omega = \pm\omega_{bp} - i\frac{\nu}{2}\frac{\omega_p^2}{\omega_{bp}^2},\ \omega_{bp} = \sqrt{\omega_p^2+\lambda^2c^2}\,,$$

which are the roots of the equation

$$k_z(\omega) = \frac{1}{c}\sqrt{\omega^2\varepsilon(\omega)-\lambda^2c^2} = 0$$

and lie in the lower half-plane of the complex variable $\omega$. Let's make cuts $\mathrm{Im}\,k_z = 0$ on the complex plane (see Fig. 1). On the upper sheet of the Riemann surface, the real part $k_z(\omega)$ on the upper and lower sides of the cut has opposite signs. Note that in integrand (24) for the magnetic field, poles (14) corresponding to Langmuir oscillations are absent.

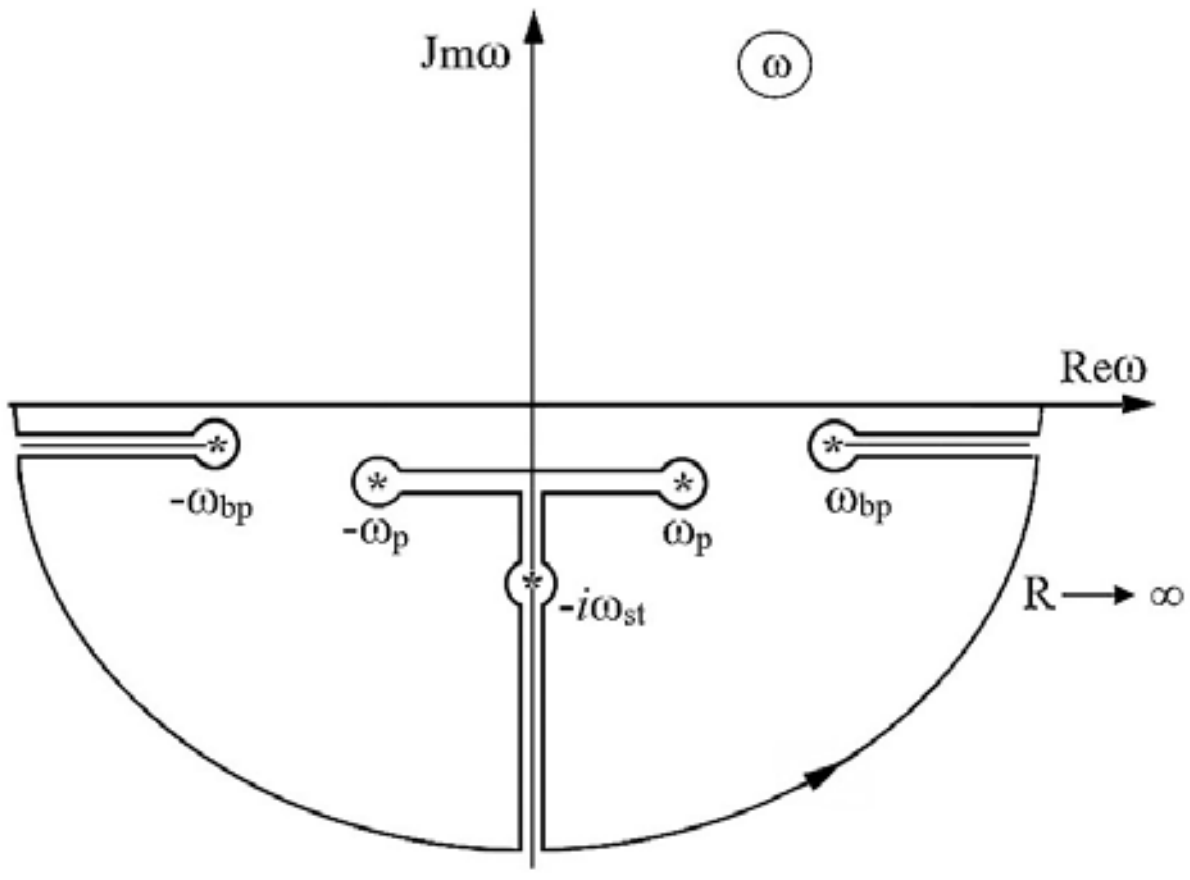

Fig.1. Integration contour

In the upper half-plane of the complex variable $\omega$, the integrand in (22) has only one simple pole $\omega = i\omega_{st}$. This pole corresponds to the quasi-static (quasi-Coulomb) field of the electron bunch. When

$$z > c\bar{t} \qquad (25)$$

it is necessary to close the integration contour into the upper half-plane of the complex variable $\omega$. Calculating the residue in the pole $\omega = i\omega_{st}$, we find the value of the Fourier integral (21) under the condition (25)

$$E_{GTr}(z > c\bar{t}) = -E_{GBr}\,.$$

Accordingly, the full field behind the wave front $z > c\bar{t}$ is equal zero

$$\vec{E}_G(z > c\bar{t}) = 0\,.$$

The result is obvious, since electromagnetic radiation simply has not yet arrived to this region.

Let`s consider now the region

$$z < c\bar{t}\,.$$

In this region, in the Fourier integral (22) the integration contour must be closed in the lower half-plane of the complex variable $\omega$. The integration contour $C$ is shown in Fig. 1. In the inner region bounded by the contour, the integrand in the integral (22) has no peculiarities. Therefore, the integral along this contour is equal to zero. As a result, we obtain that the sought Fourier integral (22) is equal to the sum of the residues in the poles (14) corresponding to the Langmuir oscillations, the residue in the pole $\omega = -i\omega_{st}$ responsible for the quasi-Coulomb field of the electron bunch, plus the integral over the loop $C_{loop}$ that encloses the cut

$$F_{Tr}(\bar{t},z,\lambda) = -2\pi i\Big[resF_{Tr}(\bar{t},z,\lambda,\omega_p) + resF_{Tr}(\bar{t},z,\lambda,-\omega_p)$$
$$+ resF_{Tr}(\bar{t},z,\lambda,-i\omega_{st})\Big] + \int_{C_{loop}} F_{Tr\omega}(z,\lambda)d\omega + c.c.\,. \quad (26)$$

The integral over the cut in the left half-plane of the complex variable $\omega$ ($\mathrm{Re}\,\omega < 0$) is complex conjugate.

For the magnetic field component $H_{GT\varphi}$ we have the similar expression

$$F_{T\varphi}(\overline{t},z,\lambda) = -2\pi i res F_{T\varphi}(\overline{t},z,\lambda,-i\omega_{st}) + + \int\limits_{C_{loop}} F_{T\varphi\omega}(z,\lambda)d\omega + c.c. \quad (27)$$

with the only difference that there are no residues in the poles corresponding to Langmuir oscillations.

After calculation the residues in the poles $\omega = \pm\omega_p - i\nu/2$, we find the transition field of Langmuir oscillations

$$\Phi_{GT} = -2dQk_p G_0(\eta,\eta_0,\varsigma)\vartheta(\tau-\tau_{Lem})\sin(\tau-\tau_0)\big],$$

where $\tau_{Lem} = \omega_p(t_0 + z/c)$ describes the motion of the front of the electromagnetic field

$$G_0(\eta,\eta_0,\varsigma) = \int\limits_0^\infty J_0(\lambda\eta_0)J_0(\lambda\eta)\frac{e^{-\lambda\varsigma}}{\lambda^2+1}\lambda d\lambda, \quad (28)$$

Accordingly, for the full electric potential of the Langmuir oscillations, we have the expression

$$\Phi_G = \Phi_{Gw} + \Phi_{GT} = = 2dQk_p\left[\Gamma_0(\eta,\eta_0)\vartheta(\tau-\tau_L)\sin(\tau-\tau_L) - -G_0(\eta,\eta_0,\varsigma)\vartheta(\tau-\tau_{Lem})\sin(\tau-\tau_0)\right]. \quad (29)$$

It is easy to verify that the electric potential $\Phi_{Gw}$ satisfies the boundary condition

$$\Phi_G(\eta,\eta_0,\tau,\varsigma=0) = 0.$$

Thus, in a semi-bounded plasma, the field of Langmuir oscillations excited by an electron bunch contains a monochromatic wake wave and a clot of Langmuir oscillations localized near the plasma boundary. The function $G_0(\eta,\eta_0,\varsigma)$ describes the spatial distribution of the electric potential of the transition field of Langmuir oscillations.

For a point bunch $\eta_0 = 0$, the expression for function (28) is simplified

$$G_0(\eta,0,\varsigma) = \int\limits_0^\infty J_0(\lambda\eta)\frac{e^{-\lambda\varsigma}}{\lambda^2+1}\lambda d\lambda.$$

At large distances from the plasma boundary $\varsigma >> 1$, the main contribution to this integral comes from the region of small values of the variable $\lambda << 1$. In this limiting case, we obtain an approximate value of this integral

$$G_0(\eta,0,\varsigma) = \frac{\varsigma}{\left(\varsigma^2+\eta^2\right)^{3/2}}. \quad (30)$$

In a spherical coordinate system, instead of (30), we have

$$G_0(\eta,0,\varsigma) = \frac{\cos\vartheta}{\rho^2}, \quad \rho = \sqrt{\varsigma^2+\eta^2}.$$

Let us now consider the structure of the quasi-static transition electromagnetic field of a relativistic electron bunch moving in plasma. The residue in the pole $\omega = -i\omega_{st}$ gives an expression for all components of the transition quasi-static electromagnetic field of the electron bunch

$$E_{GTr}^{(st)} = dQk_p^2\gamma_0^2\int\limits_0^\infty \Pi_1(\lambda\eta)\frac{\lambda^2\kappa e^{-\kappa(\lambda)|\varsigma+\varsigma_L|}}{\kappa^2(\lambda)+1}d\lambda, \quad (31)$$

$$E_{GTz}^{(st)} = dQk_p^2\gamma_0^2\int\limits_0^\infty \Pi_0(\lambda\eta)\frac{\lambda^3 e^{-\kappa(\lambda)|\varsigma+\varsigma_L|}}{\kappa^2(\lambda)+1}d\lambda, \quad (32)$$

$$H_{GT\varphi}^{(st)} = -dQk_p^2\beta_0\gamma_0^2\int\limits_0^\infty \Pi_1(\lambda\eta)\frac{\lambda^2 e^{-\kappa(\lambda)|\varsigma+\varsigma_L|}}{\kappa(\lambda)}d\lambda, \quad (33)$$

$\beta_0 = v_0/c$.

These expressions describe the field of a positive charge induced by an electron bunch at an perfectly conducting plasma boundary (image field). The full quasi-static field of a moving electron bunch will consist of the sum of the fields of the charge itself and its oppositely charged image

$$\vec{E}_G^{(st)}(\eta,\varsigma) = \vec{E}_{GB}(\eta,|\varsigma-\varsigma_L(\tau,\tau_0)|) + \vec{E}_{GT}(\eta,\varsigma+\varsigma_L(\tau,\tau_0)),$$

$$\vec{H}_G^{(st)}(\eta,\varsigma) = \vec{H}_{GB}(\eta,|\varsigma-\varsigma_L(\tau,\tau_0)|) + \vec{H}_{GT}(\eta,\varsigma+\varsigma_L(\tau,\tau_0)).$$

At the perfectly conducting plasma boundary $z = 0$, the radial component of the full quasi-static electric field vanishes.

For a point electron bunch $\eta_0 = 0$, integrals (19), (33) describing the magnetic field of the bunch and its image are calculated exactly [10]. As a result, the expressions for the total quasi-static magnetic field of a point electron bunch can be represented as

$$H_{G\varphi}^{(st)} = -Qk_p^2\beta_0\gamma_0\left[\frac{\eta}{\rho_L^{(-)^3}}\left(1+\beta_0\rho_L^{(-)}\right)e^{-\beta_0\rho_L^{(-)}} + +\frac{\eta}{\rho_L^{(+)^3}}\left(1+\beta_0\rho_L^{(+)}\right)e^{-\beta_0\rho_L^{(+)}}\right], \quad (34)$$

$$\rho_L^{(\pm)} = \sqrt{\eta^2+\gamma_0^2\left(\varsigma\pm\varsigma_L\right)^2}.$$

For such a bunch, the elementary charge $dQ$ of the bunch should be understood as the total charge $Q$. In the ultrarelativistic case $\gamma_0 >> 1$, up to the magnitudes of the order $\gamma_0^{-2}$, the transverse components of the electric and magnetic fields (18), (19) of the point bunch coincide. As for the transverse fields of the image (31), (33) , they are close in magnitude, but their signs are opposite. In this limiting case, the expression for the radial component of the total quasi-static electric field has the form

$$E_{Gr}^{(st)} = -Qk_p^2\gamma_0\left[\frac{\eta}{\rho_L^{(-)^3}}\left(1+\beta_0\rho_L^{(-)}\right)e^{-\beta_0\rho_L^{(-)}} - -\frac{\eta}{\rho_L^{(+)^3}}\left(1+\beta_0\rho_L^{(+)}\right)e^{-\beta_0\rho_L^{(+)}}\right]. \quad (35)$$

It follows from relations (34) and (35) that the quasi static field decreases exponentially with distance from the bunch. The physical reason for the exponential decreasing the quasi-static electromagnetic field is its screening by plasma. In the transverse direction, the quasi-static electromagnetic field decays at the depth of the skin layer $c/\omega_p$ and in the longitudinal direction on scale $c/(\omega_p\gamma_0)$, i.e. inversely proportional to the relativistic factor.

Let us proceed to the analysis of the last integral term in the expression (26), which describes the electromagnetic field of transition radiation. Taking into account that the signs are opposite on the upper and lower sides of the cut $\operatorname{Im} k_z = 0$ (see Fig. 1), the integral (26) over the loop $C_{loop}$ can be transformed to the form

$$F_{Tr}^{(rad)}(t,z,\lambda) = 2i\int_{\omega_{bp}}^{\infty} f_{Tr}^{(rad)}(\omega,z,\lambda)e^{-i\omega\bar{t}}d\omega + c.c.,$$

$$f_{Tr}^{(rad)}(\omega,z,\lambda) = \frac{\sin k_z z}{\varepsilon(\omega)\left(k_z^2 - k_l^2\right)}.$$

Accordingly, the expression for the radial component of the radiation field we have the expression

$$E_{GTr}^{(rad)} = -i\frac{2dQ}{\pi v_0}\int_0^{\infty}\Pi_1(\lambda r)\lambda^2 d\lambda \times$$

$$\times\int_{\omega_{bp}}^{\infty} f_{Tr}^{(rad)}(\omega,z,\lambda)e^{-i\omega\bar{t}}d\omega + c.c.. \quad (36)$$

Note that the radial component of the transition electromagnetic field (36) vanishes on a perfectly conductive surface $z = 0$. For further analysis, it is convenient to change the order of integration in integral (36). As a result, we obtain

$$E_{GTr}^{(rad)} = -\frac{2idQ}{\pi v_0}\int_{\omega_p}^{\infty}\frac{e^{-i\omega\bar{t}}}{\varepsilon(\omega)}I_E(\omega)d\omega + c.c., \quad (37)$$

$$I_E(\omega) = \int_0^{k_0\sqrt{\varepsilon}} P(\lambda,\omega)J_1(\lambda r)\sin k_z z d\lambda, \quad (38)$$

$$P(\lambda,\omega) = J_0(\lambda r_0)\frac{\lambda^2}{\lambda^2 + k_l^2 - k_0^2\varepsilon(\omega)}.$$

We also present an expression for the magnetic field

$$H_{GT\varphi}^{(rad)} = -i\frac{2dQ}{\pi v_0}\int_{\omega_p}^{\infty} k_0 e^{-i\omega\bar{t}} I_H(\omega)d\omega + c.c., \quad (39)$$

$$I_H(\omega) = \int_0^{k_0\sqrt{\varepsilon}}\frac{P(\lambda,\omega)}{k_z(\lambda,\omega)}J_1(\lambda r)\cos k_z z d\lambda. \quad (40)$$

The next step of the transformation will be the change of the variable $\lambda = k_0\sqrt{\varepsilon}\sin w$ and the transition to the spherical coordinate system $z = R\cos\vartheta,\ r = R\sin\vartheta$. Then, instead of integral (38), we obtain

$$I_E(\omega) = \frac{\left(k_0\sqrt{\varepsilon}\right)^3}{2ik_l^2}\left[\int_0^{\pi/2} P(w)J_1(\rho_\omega\sin\vartheta\sin w)e^{i\rho_\omega\cos\vartheta\cos w}dw - \right.$$

$$\left. -\int_0^{\pi/2} P(w)J_1(\rho_\omega\sin\vartheta\sin w)e^{-i\rho_\omega\cos\vartheta\cos w}dw\right], \quad (41)$$

where

$$P(w) = J_0(\rho_0\sin w)\frac{\sin^2 w\cos w}{1 - \beta_0^2\varepsilon(\omega)\cos^2 w},$$

$$\rho_\omega = k_0\sqrt{\varepsilon(\omega)}R,\ \rho_0 = k_0\sqrt{\varepsilon(\omega)}r_0.$$

In the wave zone $\rho_\omega >> 1$, we can use the asymptotic representation of the Bessel functions for large values of the argument

$$J_n(z) = \sqrt{\frac{2}{\pi z}}\cos\left(z - \frac{n\pi}{2} - \frac{\pi}{4}\right).$$

Integrals (41) in this limiting case are simplified and take the form

$$I_E(\omega) = \frac{\left(k_0\sqrt{\varepsilon}\right)^3}{4ik_l^2}\sqrt{\frac{2}{\pi k_0\sqrt{\varepsilon}R\sin\vartheta}}\left[e^{-i\frac{3\pi}{4}}S^+(\omega) + \right.$$

$$\left. + e^{i\frac{3\pi}{4}}S^-(\omega)\right] + c.c.. \quad (42)$$

Here

$$S^{\pm}(\omega) = \int_0^{\pi/2}\frac{P(w)}{\sqrt{\sin w}}e^{i\rho_\omega\cos(w\mp\vartheta)}dw. \quad (43)$$

For large $\rho_\omega$ (in the wave zone $\rho_\omega >> 1$), the exponents in $S^{\pm}(\omega)$ the integrands fast oscillate and these oscillations compensate each other in most of the area of integration. The exceptions are end points $w = 0;\pi/2$ and also points of the stationary phase, in the vicinity of which the phase functions $\varphi^{\pm}(w) = \cos(w\mp\vartheta)$ in the integrals $S^{\pm}(\omega)$ change slowly. The exponent in the integrand function of the integral $S^+(\omega)$ has a stationary phase point $w = \vartheta$, which is located within the limits of integration. Therefore, for an asymptotic estimate of this integral, one can use the stationary phase method [18-20]. Application of this method, as well as taking into account the contribution of end points, gives the following approximate expression for the function $S^+(\omega)$

$$S^+(\omega) = P(\vartheta)\sqrt{\frac{2\pi}{k_0\sqrt{\varepsilon}R\sin\vartheta}}e^{i(k_0\sqrt{\varepsilon}R - \pi/4)} +$$

$$+S^+(0) + S^+(\pi/2), \quad (44)$$

$$S^+(0) = -\frac{3\sqrt{\pi}}{4\left(1-\beta_0^2\varepsilon\right)}\frac{e^{i\left(k_0\sqrt{\varepsilon}R\cos\vartheta + \pi/4\right)}}{\left(k_0\sqrt{\varepsilon}R\sin\vartheta\right)^{5/2}}, \quad (45)$$

$$S^+(\pi/2) = -J_0(k_0\sqrt{\varepsilon}r_0)\frac{e^{ik_0\sqrt{\varepsilon}R\sin\vartheta}}{\left(k_0\sqrt{\varepsilon}R\cos\vartheta\right)^2}. \quad (46)$$

The first term in expression (44) takes into account the stationary point, and the second and third terms take account the end points $w = 0;\ \pi/2$, respectively.

Let us now turn to the integral $S^-(\omega)$. The exponential integrand of this integral has a stationary phase point that is outside the limits of integration and does not contribute to the integral. Under these conditions, the main contribution to this integral is made only by the end points of the area of integration

$$S^-(\omega) = S^-(0) + S^-(\pi/2). \quad (47)$$

$$S^-(0) = -\frac{3\sqrt{\pi}}{4\left(1-\beta_0^2\varepsilon\right)}\frac{e^{i\left(k_0\sqrt{\varepsilon}R\cos\vartheta - \pi/4\right)}}{\left(k_0\sqrt{\varepsilon}R\sin\vartheta\right)^{5/2}},$$

$$S^-(\pi/2) = -J_0(k_0\sqrt{\varepsilon}r_0)\frac{e^{-ik_0\sqrt{\varepsilon}R\sin\vartheta}}{\left(k_0\sqrt{\varepsilon}R\cos\vartheta\right)^2}. \quad (48)$$

It is easy to make sure that the contribution of end points (45), (46), (47), (48) to expression (42) fully compensates each other. As a result, the expression for the function $I_E(\omega)$ takes the form

$$I_E(\omega) = -\frac{\beta_0^2 \varepsilon \sin\vartheta \cos\vartheta}{1-\beta_0^2 \varepsilon \cos^2\vartheta} J_0\left(\rho_0 \sin\vartheta\right) \frac{\sin\left(k_0 \sqrt{\varepsilon} R\right)}{R} . \quad (49)$$

Accordingly, for the radial component of the electric field of electromagnetic radiation (37), we obtain the following integral Fourier representation

$$E_{GTr}^{(rad)} = \frac{2dQ}{\pi c} \beta_0 \frac{\sin\vartheta \cos\vartheta}{R} \left[ \Gamma^+(R,\bar{t}) - \Gamma^-(R,\bar{t}) \right], \quad (50)$$

$$\Gamma^{\pm}(R,\bar{t}) = \int_{\omega_p}^{\infty} M(\omega,\vartheta) \cos\left( \omega\bar{t} \mp \frac{R}{c}\sqrt{\omega^2 - \omega_p^2} \right) d\omega , \quad (51)$$

$$M(\omega,\vartheta) = \frac{J_0\left(\rho_0 \sin\vartheta\right)}{1-\beta_0^2 \varepsilon(\omega) \cos^2\vartheta} .$$

Let us make change $\kappa = k_z(\omega) \equiv \frac{1}{c}\sqrt{\omega^2 - \omega_p^2}$ in integrals (51). As the independent variable we have chosen the longitudinal wavenumber. In terms of the new variable, integrals (51) take the form

$$\Gamma^{\pm}(R,\bar{t}) = c\int_0^{\infty} \Gamma(\kappa) \cos\left(\bar{t}\psi^{(\pm)}(\kappa)\right) d\kappa, \quad (52)$$

where

$$\Gamma(\kappa) = \frac{\kappa J_0\left(\kappa r_0 \sin\vartheta\right)}{\left(1-\beta_0^2 \varepsilon(\kappa)\cos^2\vartheta\right)\sqrt{\kappa^2 + \kappa_p^2}} ,$$

$$\varepsilon(\kappa) = 1 - \frac{\omega_p^2}{\omega^2(\kappa)} \equiv \frac{\kappa^2}{\kappa^2 + \kappa_p^2}, \; \omega(\kappa) = \sqrt{\omega_p^2 + \kappa^2 c^2} ,$$

$$\kappa_p = \frac{\omega_p}{c}, \; \psi^{(\pm)}(\kappa) = \omega(\kappa) \mp \kappa u, \; u = R/\bar{t} .$$

Let's consider the asymptotic representations of integrals (52) for large $\bar{t}$. To estimate these integrals, we again turn to the stationary phase method. The points of the stationary phases are the roots of the equations

$$\frac{d\psi^{(\pm)}(\kappa)}{d\kappa} = 0 . \quad (53)$$

The expressions for these roots are as follows

$$\kappa = \kappa_s^{(\pm)}(\bar{\varsigma}) \equiv \pm \frac{\kappa_p \bar{\varsigma}}{\sqrt{1-\bar{\varsigma}^2}}, \; \bar{\varsigma} = \frac{u}{c} \le 1 . \quad (54)$$

Formal equation (53) is equivalent to the physically more visual one

$$v_g(\kappa) = \frac{d\omega(\kappa)}{d\kappa} = \pm\frac{R}{\bar{t}} \equiv \pm c\bar{\varsigma} .$$

The group velocity signs $v_g(\omega)$ correspond to wave packets propagating in opposite directions. The stationary point $\kappa = \kappa_s^{(+)}(\bar{\varsigma})$ of the phase function $\psi^{(+)}(\kappa)$ is within the limits of integration of the integral $\Gamma^+(R,\bar{t})$ and it must be taken into account. As for the stationary point $\kappa = \kappa_s^{(-)}(\bar{\varsigma})$, it is outside the limits of integration of the integral $\Gamma^-(R,\bar{t})$ and does not make a contribution. In the vicinity of the stationary point $\kappa = \kappa_s^{(+)}(\bar{\varsigma})$, the phase function $\psi^{(+)}(\kappa)$ can be expanded in the series

$$\psi^{(+)}(\kappa) = \psi^{(+)}(\kappa_s^{(+)}) + \frac{1}{2}\omega''(\kappa_s^{(+)})\left(\kappa - \kappa_s^{(+)}\right)^2 ,$$

where

$$\omega''(\kappa_s^{(+)}(\bar{\varsigma})) = \frac{c^2}{\omega_p}\left(1-\bar{\varsigma}^2\right)^{3/2} ,$$

$$\psi^{(+)}(\kappa_s^{(+)}) = \omega(\kappa_s^{(+)}(\bar{\varsigma})) - \kappa_s^{(+)}(\bar{\varsigma}) c\bar{\varsigma} .$$

The dependence of the longitudinal wave number $\kappa_s^{(+)}$ on the value $\bar{\varsigma}$ is determined by expression (54). For the frequency $\omega_s(\bar{\varsigma}) \equiv \omega(\kappa_s^{(+)}(\bar{\varsigma}))$ and dielectric constant $\varepsilon_s\left(\bar{\varsigma}\right) \equiv \varepsilon\left(\omega_s(\bar{\varsigma})\right)$, we have simple expressions

$$\omega_s(\bar{\varsigma}) = \frac{\omega_p}{\sqrt{1-\bar{\varsigma}^2}} . \; \varepsilon_s\left(\bar{\varsigma}\right) = \bar{\varsigma}^2 . \quad (55)$$

The dielectric constant of the plasma changes from zero at the plasma boundary $R = 0$ (cutoff frequency for electromagnetic waves) to unity at the spherical front of the transition pulse $R = c\bar{t}$. When conditions are met

$$\Lambda = \frac{1}{2}\bar{t}\,\omega''(\kappa_s^{(+)})\kappa_s^{(+)^2} = \frac{1}{2}\omega_p \bar{t}\bar{\varsigma}^2\sqrt{1-\bar{\varsigma}^2} >> 1 , \quad (56)$$

the main contribution to the integral $\Gamma^+(R,\bar{t})$ comes from a narrow vicinity of the stationary phase point $\kappa = \kappa_s^{(+)}(\bar{\varsigma})$. In this limiting case, we obtain the following approximate value of integral

$$\sqrt{\quad} \Gamma^+(\bar{t},R) = \Gamma_s(\bar{\varsigma},\vartheta)\sqrt{\frac{2\pi}{\bar{t}\,\omega_s''(\bar{\varsigma})}} \, e \, \mathrm{Re}^{-i\left[\omega_s(\bar{\varsigma})\bar{t} - \kappa_s^{(+)}(\bar{\varsigma})R + \frac{\pi}{4}\right]} ,$$

$$\Gamma_s(\bar{\varsigma},\vartheta) \equiv \Gamma(\kappa_s(\bar{\varsigma})) , \; \omega_s''(\bar{\varsigma}) \equiv \omega''(\kappa_s^{(+)}(\bar{\varsigma})) .$$

As noted above, the stationary phase point $\kappa = \kappa_s^{(-)}(\bar{\varsigma})$ is outside the integration limits of the integral $\Gamma^-(R,\bar{t})$. Under these conditions, at $\bar{t} >> 1$, only the end point $\kappa = 0$ will make the main contribution to the value of this integral. In principle, the end point will also make a certain contribution to the integral $\Gamma^+(R,\bar{t})$. The phase function $\psi^{(-)}(\kappa)$ and all its derivatives are continuous in the area of integration, therefore, the contribution of the end point to the asymptotic estimate of the integrals under study at $\bar{t} \to \infty$ can be obtained by integrating by parts [19]. As a result, we obtain the following expressions for the functions $\Gamma_{end}^{\pm}(R,\bar{t})$ taking into account the contribution of the end point $\kappa = 0$ to integrals (52)

$$\Gamma_{end}^{\pm}(R,\bar{t}) = -\frac{c}{\omega_p R^2} e^{-i\omega_p \bar{t}} .$$

From expression (50) for the electric field of a transition electromagnetic pulse, it is seen that the terms $\Gamma_{end}^{\pm}(R,\bar{t})$ compensate each other. As a result, for the radial component of the electric field of the pulse of electromagnetic transition radiation, we obtain the following expression

$$E_{GTr}^{(rad)}(R,\bar{t}) = \frac{2}{\pi} dQ\beta_0 \frac{\sin\vartheta\cos\vartheta}{R} \Gamma_s(\bar{\varsigma},\vartheta)\sqrt{\frac{2\pi}{\bar{t}\,\omega_s''(\bar{\varsigma})}} \times$$

$$\times \cos\left[\omega_s(\bar{\varsigma})\bar{t} - \kappa_s^{(+)}(\bar{\varsigma})R + \frac{\pi}{4}\right] , \quad (57)$$

where

$$\Gamma_s(\varsigma,\vartheta)=\frac{\bar{\varsigma}J_0\left(\kappa_s^{(+)}r_0\sin\vartheta\right)}{1-\beta_0^2\bar{\varsigma}^{-2}\cos^2\vartheta}.$$

Thus, a transition electromagnetic pulse is a wave packet with variable frequency $\omega_s(\bar{\varsigma})$ and wave number $\kappa_s^{(+)}(\bar{\varsigma})$. It follows from relations (55), (54) that the frequency $\omega_s(\bar{\varsigma})$ and the local wave number $\kappa_s^{(+)}(\bar{\varsigma})$ propagate in space (plasma) with the group velocity $v_g(\bar{\varsigma})=c\bar{\varsigma}=R/\bar{t}, v_g(\bar{\varsigma})\equiv v_g(\kappa_s^{(+)}(\bar{\varsigma}))$, i.e. conserve their value along the characteristic $R=v_g(\bar{\varsigma})\bar{t}$. Note that the group and phase velocities are related to each other by a well-known relation $v_g(\bar{\varsigma})v_{ph}(\bar{\varsigma})=c^2$. From the expression for frequency (55) and wave number (54) it also follows that the high-frequency (short-wave) component of the transition electromagnetic pulse is concentrated in the region of the leading edge of the transition pulse $R\le c\bar{t}$. With distance from the leading edge to the plasma boundary, the frequency decreases, and the wavelength, accordingly, increases.

Using expressions for $\omega_s(\bar{\varsigma})$, $\kappa_s^{(+)}(\bar{\varsigma})$ and, $\omega_s''(\bar{\varsigma})$ relation (57) for the amplitude of the field of an electromagnetic pulse can be written in the form

$$E_{GTr}^{(rad)}(R,\vartheta,\bar{t})=2dQ\kappa_p\beta_0\frac{\sin\vartheta\cos\vartheta}{R}\frac{f(\bar{\varsigma})}{\bar{\varsigma}}J_0\left(\kappa_s^{(+)}r_0\sin\vartheta\right)\times$$
$$\times\sqrt{\frac{2}{\pi\omega_p\bar{t}}}\cos\left(\omega_p\bar{t}\sqrt{1-\bar{\varsigma}^2}+\frac{\pi}{4}\right), \qquad (58)$$

where

$$f(\bar{\varsigma})=\frac{\bar{\varsigma}^2}{\left(1-\beta_0^2\bar{\varsigma}^{-2}\cos^2\vartheta\right)\left(1-\bar{\varsigma}^2\right)^{3/4}}.$$

We also present an expression for the magnetic field of the transition pulse

$$H_{GT\varphi}^{(rad)}(R,\vartheta,\bar{t})=2dQ\kappa_p\beta_0\frac{\sin\vartheta}{R}f(\bar{\varsigma})J_0\left(\kappa_s^{(+)}r_0\sin\vartheta\right)\times$$
$$\times\sqrt{\frac{2}{\pi\omega_p\bar{t}}}\cos\left(\omega_p\bar{t}\sqrt{1-\bar{\varsigma}^2}+\frac{\pi}{4}\right). \qquad (59)$$

When approaching to the leading edge of the transition pulse $R\to c\bar{t}$, the field amplitude increases, and at the leading edge $R=c\bar{t}$ the field has an integrable singularity. To understand the nature of this feature, one should pay attention to the fact that in the high-frequency region of the spectrum $\omega>>\omega_p$, the Fourier frequency amplitude (51) weakly depends on the frequency

$$M(\omega>>\omega_p,\vartheta)=\frac{1}{1-\beta_0^2\cos^2\vartheta}J_0\left(\frac{\omega r_0}{c}\sin\vartheta\right).$$

At the leading edge of the transition pulse, all high-frequency the Fourier harmonics of the field of the transition pulse are concentrated in the vicinity of the leading edge of the pulse and are coherently added, forming a singularity of the field in the frame of the approximation of the stationary phase method. Meanwhile, when approaching to the leading edge of the transition pulse, the value of parameter $\Lambda$ (56) also decreases and at $|\Lambda|\le 1$, i.e. in the vicinity of the leading edge of the transition pulse, the stationary phase method becomes incorrect. To determine the field in the vicinity of the leading edge of the transition pulse $R=c\bar{t}$, we represent integral (52) $\Gamma^+(\bar{t})\equiv\Gamma^+(R=c\bar{t},\bar{t})$ in the form

$$\Gamma^+(\bar{t})=c\int_0^\infty\frac{J_0\left(\kappa r_0\sin\vartheta\right)}{\left(1-\beta_0^2\varepsilon(\kappa)\cos^2\vartheta\right)}\frac{\kappa\cos\left[c\bar{t}\phi(\kappa)\right]}{\sqrt{\kappa^2+\kappa_p^2}}d\kappa, \qquad (60)$$

where $\phi(\kappa)=\sqrt{\kappa_p^2+\kappa^2}-\kappa$. The value of the function $\phi(\kappa)$ is in the limits $\kappa_p>\phi(\kappa)>0$ and decreases monotonically to zero at $\kappa\to\infty$. To estimate from above the integral (60), we can replace the included in the integrand $\cos\left[c\bar{t}\phi(\kappa)\right]$ (60) by unity. As a result, we obtain

$$\Gamma^+(\bar{t})=c\int_0^\infty\frac{J_0\left(\kappa r_0\sin\vartheta\right)}{\left(1-\beta_0^2\varepsilon(\kappa)\cos^2\vartheta\right)}\frac{\kappa}{\sqrt{\kappa^2+\kappa_p^2}}d\kappa. \qquad (61)$$

For a point bunch $r_0=0$, integrals (60), (61) diverge, i.e., the field singularity is preserved at the leading edge of the transition pulse. However, for a ring bunch, the integral has a finite value. The region $\kappa>>1$ will make the main contribution to the integral. Taking this circumstance into account, the integrand (61) can be replaced by its asymptote at $\kappa>>1$. As a result, we obtain an upper estimate for the value of the function

$$\Gamma^+=\frac{c}{1-\beta_0^2\cos^2\vartheta}\int_0^\infty J_0\left(\kappa r_0\sin\vartheta\right)d\kappa=$$
$$=\frac{c}{r_0\sin\vartheta\left(1-\beta_0^2\cos^2\vartheta\right)}$$

and, accordingly, the fields at the leading edge of the transition pulse

$$E_{GTr}^{(rad)}=\frac{2dQ}{\pi r_0}\frac{\beta_0\cos\vartheta}{1-\beta_0^2\cos^2\vartheta}\frac{1}{R}.$$

For a point bunch $r_0=0$, the singularity of the field at the leading edge of the transition pulse always takes place, however, for a ring bunch, the interference of the Fourier amplitudes of the radiated field leads to the elimination of the singularity and the establishment of a finite field value at the leading edge of the transition pulse.

## 3. EXCITATION OF TRANSITION ELECTROMAGNETIC FIELDS BY RELATIVISTIC ELECTRON BUNCH OF FINITE SIZES

It was shown above that in the plasma half-space an infinitely thin ring relativistic electron bunch excites Langmuir oscillations at the electron plasma frequency, an own quasi-static electromagnetic field and the field of the positively charged image, as well as the transition electromagnetic pulse.

Let us first consider the process of excitation of potential plasma oscillations by a relativistic electron bunch of finite dimensions. The electric potential of plasma oscillations excited by an elementary ring electron bunch (Green's function) is described by formula (29). Then, for an electron bunch with density profile (5), after integration over the enter times and the initial radii of elementary ring bunches, we obtain the following expression for potential of the plasma oscillations

$$\Phi(\eta,\varsigma,\tau) = \Phi_w(\eta,\hat{\tau}) + \Phi_T(\eta,\varsigma,\tau), \qquad (62)$$

where $\hat{\tau} = \omega_p (t - z / v_0)$,

$$\Phi_w = \Phi_0 \Pi_w(\eta) Z_w(\hat{\tau}) \qquad (63)$$

is electric potential of a wake plasma wave in an unbounded plasma,

$$\Pi_w(\eta) = \int_0^\infty R\left(\frac{\eta_0}{\eta_b}\right)\Gamma(\eta,\eta_0)\eta_0 d\eta_0, \qquad (64)$$

$$Z_w(\hat{\tau}) = \int_{-\infty}^{\hat{\tau}} T\left(\frac{\tau_0}{\tau_b}\right)\sin(\hat{\tau}-\tau_0)d\tau_0, \qquad (65)$$

$$\Phi_T(\eta,\tau,\varsigma) = -\Phi_0 \Pi_T(\eta,\varsigma) Z_T(\tau,\varsigma) \qquad (66)$$

is electric potential of transition plasma oscillations,

$$Z_T(\tau,\varsigma) = \int_{-\infty}^{\tau-\beta_0\varsigma} T\left(\frac{\tau_0}{\tau_b}\right)\sin(\tau-\tau_0)d\tau_0, \qquad (67)$$

$$\Pi_T(\eta,\varsigma) = \int_0^\infty J_0(\lambda\eta)Y(\lambda)\frac{e^{-\lambda\varsigma}}{\lambda^2+1}\lambda d\lambda, \qquad (68)$$

$$Y(\lambda) = \int_0^\infty R\left(\frac{\eta_0}{\eta_b}\right)J_0(\lambda\eta_0)\eta_0 d\eta_0, \qquad (69)$$

$$\Phi_0 = \frac{4\pi Q k_p}{\sigma_{eff}\tau_{eff}},\ \sigma_{eff} = s_{eff}k_p^2, \tau_{eff} = \omega_p t_{eff},\ \eta_b = k_p r_b,$$

$\tau_b = \omega_p t_b$.

The transition electric potential of plasma oscillations is localized in the vicinity of the conducting wall. It is easy to verify that at the perfectly conducting boundary of the plasma half-space, electric potential (62) vanishes

$$\Phi(\varsigma = 0,\tau) = 0. \qquad (70)$$

The wake function $Z_w(\hat{\tau})$ describes the distribution of the wake field in the longitudinal direction at each moment of time. We will consider an electron bunch with a symmetric longitudinal profile $T(\tau_0/\tau_b) = T(-\tau_0/\tau_b)$. The wake function $Z_w(\hat{\tau})$ is conveniently represented as

$$Z_w(\hat{\tau}) = \hat{T}\vartheta(\hat{\tau})\sin\hat{\tau} - X_w(\hat{\tau}), \qquad (71)$$

$$\hat{T} = 2\int_0^\infty T\left(\frac{\tau_0}{\tau_b}\right)\cos\tau_0 ds. \qquad (72)$$

$$X_w(\hat{\tau}) = \int_{|\hat{\tau}|}^\infty T\left(\frac{\tau_0}{\tau_b}\right)\sin(|\hat{\tau}|-\tau_0)d\tau_0 .$$

The first term in (71) describes the wake wave propagating behind the electron bunch. The amplitude of the wake wave is equal to the amplitude of the Fourier component at the plasma frequency $\hat{T}$ of the function $T(t_0/t_b)$, describing the longitudinal profile of the bunch. The second term in (71) describes a bipolar antisymmetric pulse of a polarization field localized in the region of a electron bunch. The field of this pulse decreases and tends to zero with increasing distance from the electron bunch. At large distances behind a bunch $\hat{\tau} \gg \tau_b$ (in the wave zone) or in a region of space

$$\varsigma \ll \tau - \tau_b \qquad (73)$$

for the field of wake plasma oscillations, we obtain

$$\Phi(\eta,\varsigma,\tau) = \Phi_0\hat{T}\left[\Pi_w(\eta)\sin(\tau-\varsigma) - \Pi_T(\eta,\varsigma)\sin\tau\right]. \qquad (74)$$

Thus, the steady wake field of an electron bunch contains a monochromatic plasma wave that propagates behind a relativistic electron bunch, and a transition field of plasma oscillations localized in the vicinity of the plasma boundary. At large distances from the region of injection of the electron bunch $\varsigma \gg 1, \eta \gg 1$, the law of decrease of the transition field of Langmuir oscillations (function $\Pi_T(\eta,\varsigma)$) can be determined approximately. Indeed, in this limiting case, the main contribution to integral (68) comes from the region of small values of the integration variable $\lambda \ll 1$. As a result, for the function $\Pi_T(\eta,\varsigma)$ we obtain the following approximate expression

$$\Pi_T(\eta,\varsigma) = \frac{\hat{\sigma}}{2}\eta_b^2 G_0(\eta,0,\varsigma),$$

where the function $G_0(\eta,0,\varsigma)$ is defined by expression (30). At large distances, the decrease law of the transition field excited by a bunch of finite dimensions is the same as for a point bunch.

Let us consider for definiteness an electron bunch with Gaussian longitudinal and transverse profiles

$$T\left(\frac{\tau_0}{\tau_b}\right) = \exp\left(-\frac{\tau_0^2}{\tau_b^2}\right),\ R\left(\frac{\eta_0}{\eta_b}\right) = \exp\left(-\frac{\eta_0^2}{\eta_b^2}\right). \qquad (75)$$

For such a bunch we have

$$\hat{T} = \sqrt{\pi}\tau_b e^{-\frac{\tau_b^2}{4}},\ \hat{\tau} = \sqrt{\pi},$$

$$\Pi_T(\eta,\varsigma) = \frac{\eta_b^2}{2}\int_0^\infty J_0(\lambda r)e^{-\frac{\eta_b^2\lambda^2}{4}-\lambda\varsigma}\frac{\lambda d\lambda}{\lambda^2+1}.$$

Let us turn to the study of a transition electromagnetic pulse excited by a relativistic electron bunch. Using Green's function (59), as well as relation (7), we obtain the following expression describing the shape of the magnetic field of the radiated transition pulse for an electron bunch with arbitrary longitudinal and transverse profiles

$$H_\varphi^{(rad)} = 2Q\kappa_p\beta_0\frac{\sin\vartheta}{R}Z_T(t,\varsigma,\vartheta), \qquad (76)$$

where

$$Z_T(t,\varsigma) = \frac{1}{\hat{\tau}t_b}\int_{-\infty}^{t} dt_0 T\left(\frac{t_0}{t_b}\right)\sqrt{\frac{2}{\pi\omega_p(t-t_0)}}f(\overline{\varsigma})Y_T(\overline{\varsigma})\times$$
$$\times\cos\left(\Theta(t-t_0,R)+\frac{\pi}{4}\right), \qquad (77)$$

$$\Theta(t-t_0,R) = \omega_p\sqrt{(t-t_0)^2 - \frac{R^2}{c^2}},$$

$$Y_T(\overline{\varsigma},\vartheta) = \frac{2}{\hat{\sigma}}\int_0^\infty \rho_0 d\rho_0 R(\rho_0)J_0\left(\rho_0\kappa_s^{(+)}(\overline{\varsigma})r_b\sin\vartheta\right).$$

We will consider the moments of time $t \gg t_b$ when the bunch is completely in plasma. In this case, the upper limit of integral (77) can be replaced by $+\infty$. In the case of a short electron bunch with a smooth longitudinal profile, for example, Gaussian (75), the main contribution to the integral $Z_T(z,t)$ will come from the vicinity of the point $t_0 = 0$. This allows us to expand the phase function $\Theta(t-t_0,R)$ in the vicinity of this point

$$\Theta(t-t_0,R) \approx \Theta(t,R) - \frac{t_0 t \omega_p^2}{\Theta(t,R)},$$

$$\Theta(t,R) = \omega_p \sqrt{t^2 - R^2/c^2}.$$

As a result, for integral (77) we obtain the asymptotic representation

$$Z_T(t,\varsigma) = \frac{1}{\tau_b \hat{\tau}} \hat{T}(\varsigma)\, f(\varsigma) Y_T(\varsigma) \sqrt{\frac{2}{\pi \omega_p t}} \cos\left(\Theta(t,R) + \frac{\pi}{4}\right),$$

where

$$\hat{T}(\varsigma) = 2\int_0^\infty T\left(\frac{\tau_0}{\tau_b}\right) \cos\left[\Omega_s(\varsigma)\tau_0\right] d\tau_0$$

is Fourier amplitude of the function $T(\tau_0/\tau_b)$ on the dimensionless frequency $\Omega_s(\varsigma) = \omega_s(\varsigma)/\omega_p$.

Accordingly, for the magnetic component of the transition electromagnetic pulse, instead of (76), we have

$$H_\varphi^{(rad)} = 2Q\kappa_p \beta_0 \frac{1}{\tau_b \hat{\tau}} \frac{\sin\vartheta}{R} \sqrt{\frac{2}{\pi\omega_p t}} f(\varsigma) Y_T(\varsigma,\vartheta) \hat{T}(\varsigma) \times$$

$$\times \cos\left(\Theta(t,R) + \frac{\pi}{4}\right). \quad (78)$$

We also present a simple expression for the tangential component of the electric field of the transition pulse in a spherical coordinate system.

$$E_\vartheta^{(rad)} = \frac{1}{\varsigma} H_\varphi^{(rad)} \quad (79)$$

It follows from formula (78) that the structure of the field of an electromagnetic pulse excited by a real bunch differs from the pulse of an elementary infinitely thin ring bunch by the presence of a multiplier as a form-factor $\hat{T}(\varsigma)$, the value of which, in turn, is determined by a specific longitudinal profile of the electron bunch. The form of the transition pulse strictly speaking depends on the propagation angle $\vartheta$.

For a bunch with Gaussian longitudinal transverse profiles (75), expressions for the electromagnetic field of transition radiation (78) take the form

$$H_\varphi^{(rad)} = 2Q\kappa_p \beta_0 \frac{\sin\vartheta}{R} \sqrt{\frac{2}{\pi\omega_p t}} g(\varsigma) \cos\left(\Theta + \frac{\pi}{4}\right), \quad (80)$$

$$g(\varsigma) = e^{-\frac{1}{4(1-\varsigma^2)}\left(\tau_b^2 + \rho_b^2 \varsigma^2 \sin^2\vartheta\right)} f(\varsigma).$$

Accordingly, for the electric field component $E_\vartheta^{(rad)}$ we have

$$E_\vartheta^{(rad)} = 2Q\kappa_p \beta_0 \frac{\sin\vartheta}{R} \sqrt{\frac{2}{\pi\omega_p t}} \frac{g(\varsigma)}{\varsigma} \cos\left(\Theta + \frac{\pi}{4}\right). \quad (81)$$

Note that at $\tau_b >> 1$, the amplitude of the electromagnetic pulse is exponentially small. In the case of a short electron bunch $\tau_b << 1$, a coherent excitation of a transition electromagnetic pulse takes place. The above fully applies to the dependence of the amplitude of the electromagnetic pulse on the radius of the bunch. When $\rho_b \sin\vartheta >> 1$ the bunch radiates incoherently and the amplitude of the pulse in the direction at an angle $\vartheta$ is exponentially small.

The longitudinal profile of the amplitude of the transition electromagnetic pulse at each moment of time is described by the function $g(\varsigma)$. This function vanishes at the leading edge of the transition pulse $\varsigma = 1$ and in the plane of injection of the electron bunch $\varsigma = 0$ and reaches its maximum value at the point $\varsigma = \varsigma_m$ whose coordinate is the root of the equation

$$\frac{(1-\varsigma^2)(4+3\varsigma^2) + 3\nu_0^{-2}\varsigma^2}{1-\varsigma^2+\nu_0^{-2}\varsigma^2} = \frac{q_0^2\varsigma^2}{1-\varsigma^2}, \quad (82)$$

where

$$\nu_0^{-2} = 1 - \beta_0^2\cos^2\vartheta,\; q_0^2 = \tau_b^2 + \rho_b^2\sin^2\vartheta.$$

The value of the root $\varsigma_m$ is within the limits $1 > \varsigma_m > 0$. In the most interesting case of a short and thin relativistic bunch $q_0^2 << 1$ for small observation angles $\vartheta << 1$, when the condition $\nu_0^{-2} << 1$ is also satisfied, Eq. (82) can be simplified and represented in the form

$$\frac{7(1-\varsigma^2) + 3\nu_0^{-2}}{1-\varsigma^2+\nu_0^{-2}} = \frac{q_0^2}{1-\varsigma^2}.$$

The positive root of this equation is easily found

$$\varsigma_m^2 = 1 - \frac{1}{14}\left[\sqrt{\left(q_0^2 - 3\nu_0^{-2}\right)^2 + 28\nu_0^{-2}q_0^2} + q_0^2 - 3\nu_0^{-2}\right].$$

This formula can be simplified

$$\varsigma_m = \begin{cases} 1 - q_0^2/6, & \nu_0^2 q_0^2 << 3/7, \\ 1 - q_0^2/14, & \nu_0^2 q_0^2 >> 22. \end{cases}$$

In the considered case of a relativistic electron bunch $\gamma_0 >> 1$, the function $g(\varsigma)$ that determines the spatial structure of the transition pulse at each moment of time has a sharp maximum at a point $\varsigma = \varsigma_m$. The function $g(\varsigma)$ at this point reaches its maximum value

$$g_m = \begin{cases} 3^{3/4} e^{-3/4} \nu_0^2 / q_0^{3/2}, & \nu_0^2 q_0^2 << 3/7, \\ 7^{3/4} e^{-7/4} \nu_0^2 / q_0^{3/2}, & \nu_0^2 q_0^2 >> 22. \end{cases}$$

At the point of maximum field $\varsigma = \varsigma_m$ the electromagnetic pulse frequency $\omega_s(\varsigma) = \dfrac{\omega_p}{\sqrt{1-\varsigma^2}}$ takes the value

$$\omega_s(\varsigma_m) = \begin{cases} \dfrac{\omega_p\sqrt{3}}{q_0} >> \omega_p, & \nu_0^2 q_0^2 << 3/7, \\ \dfrac{\omega_p\sqrt{7}}{q_0} >> \omega_p, & \nu_0^2 q_0^2 >> 22 \end{cases}$$

and significantly exceeds the plasma frequency.

In general, the picture of the evolution of a transition electromagnetic pulse during its propagation in the plasma half-space looks as follows. A sharp maximum of the amplitude of the wave packet is located in the vicinity of the leading edge of the pulse $z_m = c\varsigma_m t$ and propagates at a velocity $c\varsigma_m$ close to the speed of light. Since the frequency of the wave packet and the group velocity decrease with distance from the leading edge, the low-frequency components of the wave packet are constantly lagging behind. As a result, in the process of propagation, the width of the wave packet increases, and its amplitude decreases as $1/\sqrt{t}$, i.e. dispersive spreading of the transition pulse takes place.

Let's consider the question about of the power and full energy of a transition electromagnetic pulse. The power of a pulse per unit solid angle $do = \sin\vartheta d\vartheta d\varphi$ is defined as the energy flux averaged over high-frequency oscillations (Poyting's vector)

$$\frac{dP(\vartheta,\varsigma,t)}{do} = \frac{c}{4\pi}\overline{E_\vartheta^{(rad)}(R,\vartheta,\varsigma,t)H_\varphi^{(rad)}(R,\vartheta,\varsigma,t)}R^2 . \quad (83)$$

The bar in (83) denotes averaging over fast oscillations by time.

Using expressions for the field components of an electromagnetic pulse (78), (79), for the radiation power density (83) we have

$$\frac{dP(\vartheta,\varsigma,t)}{do} = \frac{W_0}{t}\sin^2\vartheta\frac{f^2(\varsigma)}{\varsigma}Y^2(\varsigma,\vartheta)\hat{T}^2(\Omega_{ps}) , \quad (84)$$

where

$$W_0 = \frac{Q^2\kappa_p\beta_0^2}{\pi^2} .$$

For a bunch with Gaussian longitudinal and transverse profiles (75), formula (84) takes the form

$$\frac{dP(\vartheta,\varsigma,t)}{do} = \frac{W_0}{t}\sin^2\vartheta\frac{f^2(\varsigma)}{\varsigma}e^{-\frac{1}{2(1-\varsigma^2)}\left(\tau_b^2+\rho_b^2\varsigma^2\sin^2\vartheta\right)} .$$

The pulse energy radiated per unit solid angle is found by integrating the pulse power density (84) by time

$$\frac{dW}{do} = \int_{R/c}^{\infty}\frac{dP(\vartheta,\varsigma,t,)}{do}dt . \quad (85)$$

Integration is carried out from the time of arrival of the leading edge of the pulse at a point $R$ formally to infinity. For an axisymmetric bunch with arbitrary longitudinal and transverse profiles, taking into account formula (84), we obtain

$$\frac{dW}{do} = W_0 S(\vartheta)\sin^2\vartheta . \quad (86)$$

$$S(\vartheta) = \int_0^1 \frac{f^2(\varsigma)}{\varsigma^2}Y_T^2(\varsigma,\vartheta)\hat{T}^2(\varsigma)d\varsigma . \quad (87)$$

For a Gaussian bunch, integral (87) takes the specific form

$$S(\vartheta) = \int_0^1 \frac{\varsigma^2 d\varsigma}{\left(1-\beta_0^2\varsigma^2\cos^2\vartheta\right)^2\left(1-\varsigma^2\right)^{3/2}}e^{-\frac{\tau_b^2+\rho_b^2\varsigma^2\sin^2\vartheta}{2(1-\varsigma^2)}} . \quad (88)$$

This integral can be reduced to the following

$$S(\vartheta) = \nu_0^4 e^{-\frac{\tau_b^2}{4}}\int_0^\infty \frac{x^2\left(1+x^2\right)d\varsigma}{\left(x^2+\nu_0^2\right)^2}e^{-\frac{q_0^2x^2}{2}}dx . \quad (89)$$

The last integral can be calculated exactly. As a result, we obtain

$$S = \frac{\sqrt{\pi}}{2}\nu_0^4 e^{-\frac{\tau_b^2}{2}}\left\{\frac{\nu_0}{y}+\sqrt{\pi}\frac{1-2\nu_0^2}{\nu_0}e^{y^2}erfc(y)+ \right.$$
$$\left. +\frac{\sqrt{\pi}}{2}\frac{1-\nu_0^2}{\nu_0}y^2\frac{d}{dy}\left[\frac{e^{y^2}}{y}erfc(y)\right]\right\}, \quad (90)$$

where $y = q_0\nu_0/\sqrt{2}$,

$$erfc(y) = \frac{2}{\sqrt{\pi}}\int_y^\infty e^{-x^2}dx$$

is error function. The function $S(\vartheta)$ can be simplified in two limiting cases: $y \ll 1$ and $y \gg 1$

$$S = \frac{\sqrt{\pi}}{2}\nu_0^4 e^{-\frac{\tau_b^2}{2}}\begin{cases}\frac{\nu_0}{y}\left[1-y\left(\frac{1}{2}+\frac{1}{\nu_0^2}\right)\right], \ y \ll 1, & (91)\\ \frac{1}{2}\frac{\nu_0}{y}\frac{1}{\nu_0^2y^2}\left(1+\frac{3}{2}\frac{\nu_0^2}{y^2}\right), \ y \gg 1. & (92)\end{cases}$$

In the considered case of the relativistic electron bunch $\gamma_0^2 \gg 1$, the condition $y \ll 1$ is satisfied in the all range of angles $\pi/2 > \vartheta > 0$ only for very short bunches $\tau_b^2\gamma_0^2 \ll 1$. For bunches with the duration $\tau_b \le 1$, approximation (92) is valid for small angles $\vartheta \ll 1$, for which

$$\tau_b\gamma_0/\sqrt{1+\vartheta^2\gamma_0^2} \gg 1 .$$

In the limiting case (91) for the energy per unit solid angle (86) of the transition pulse, we have the expression

$$\frac{dW}{do} = \sqrt{\frac{\pi}{2}}W_0 D(\vartheta) , \quad (93)$$

$$D(\vartheta) = \frac{\sin^2\vartheta}{\left(1-\beta_0^2\cos^2\vartheta\right)^2\sqrt{\tau_b^2+\rho_b^2\sin^2\vartheta}} . \quad (94)$$

The function $D(\vartheta)$ defines the radiation pattern of the energy of transition pulse energy. The angle in the direction of which the radiation pattern has a maximum is determined from the equation $dD(\vartheta)/d\vartheta = 0$, which, taking into account relation (94), is equivalent to the following

$$\frac{\sin^2\vartheta+2\alpha_b^2}{2\left(\sin^2\vartheta+\alpha_b^2\right)} = \frac{2\sin^2\vartheta}{\left(\sin^2\vartheta+\beta_0^{-2}\gamma_0^{-2}\right)} ,$$

where

$$\alpha_b = \frac{\Omega_p}{\rho_b} = \frac{ct_b}{r_b} .$$

The root of this equation in the range of angles $0 \le \vartheta \le \pi/2$ is easily found

$$\vartheta = \vartheta_m \equiv \arcsin\left[\sqrt{\frac{\left(\beta_0^{-2}\gamma_0^{-2}-2\alpha_b^2\right)^2}{36}+\frac{2}{3}\beta_0^{-2}\gamma_0^{-2}\alpha_b^2} + \frac{\beta_0^{-2}\gamma_0^{-2}-2\alpha_b^2}{6}\right] . \quad (95)$$

It is obvious that the maximum in the radiation pattern at an angle (95) is formed when the argument in (95) is less than one. This requirement is fulfilled when the condition on the parameters of the electron bunch and plasma is satisfied.

$$\gamma_0 > 2\sqrt{\frac{\alpha_b^2+1}{2\alpha_b^2+3}} > 1 .$$

The expression for angle (95) can be simplified in two limiting cases and represented as

$$\vartheta_m = \begin{cases} \arcsin\left(\frac{1}{\beta_0\gamma_0}\right) \approx \frac{1}{\gamma_0}, & \alpha_b^2\beta_0^2\gamma_0^2 >> 1, \\ \arcsin\left(\frac{1}{\sqrt{3}\beta_0\gamma_0}\right) \approx \frac{1}{\sqrt{3}\gamma_0}, & \alpha_b^2\beta_0^2\gamma_0^2 << 1. \end{cases} \quad (96)$$

In both limiting cases, the position of the radiation pattern maximum is determined only by the value of the relativistic factor $\vartheta_m \sim 1/\gamma_0$. Only numerical coefficients are different.

The maximum values of the energy density of the transition radiation in the directions determined by the angles (96) are equal

$$\left(\frac{dW}{do}\right)_{\max} = \begin{cases} \frac{1}{4}\sqrt{\frac{\pi}{2}} W_0 \frac{1}{\omega_p t_b}\frac{\gamma_0^2}{\beta_0^2}, & \alpha_b^2\beta_0^2\gamma_0^2 >> 1, \\ \frac{3}{16}\sqrt{\frac{3\pi}{2}} W_0 \frac{c}{\omega_p r_b}\frac{\gamma_0^3}{\beta_0}, & \alpha_b^2\beta_0^2\gamma_0^2 << 1. \end{cases}$$

If the condition $y >> 1$ is satisfied, and also $q_0 << 1$, then for the energy density (86) we obtain the following expression

$$\frac{dW}{do} = 3\sqrt{\frac{\pi}{2}} W_0 \frac{1}{\rho_b^2} D(\vartheta),$$

where $$D(\vartheta) = \frac{\sin^2\vartheta}{\left(\alpha_b^2+\sin^2\vartheta\right)^{5/2}} . \quad (97)$$

In the considered limiting case, the radiation pattern does not depend on the relativistic factor and is determined only by the ratio between the longitudinal and transverse dimensions of the electron bunch. The The function of the radiation pattern of the transition pulse energy $D(\vartheta)$ has a maximum for the angle

$$\vartheta_m = \arcsin\left(\sqrt{\frac{2}{3}}\alpha_b\right).$$

A short electron bunch $ct_b << r_b$ radiates the maximum transition pulse energy at small angle

$$\vartheta_m = \sqrt{\frac{2}{3}}\alpha_b << 1 .$$

The condition for the applicability $y >> 1$ of the expression for the radiation pattern (92) is equivalent to the inequality

$$y_m = \frac{\gamma_0\tau_b}{\sqrt{1+\frac{2}{3}\alpha_b^2\gamma_0^2}} >> 1 ,$$

which, in turn, is fulfilled for the values of the relativistic factor in the range

$$\frac{1}{\alpha_b} >> \gamma_0 >> \frac{1}{\tau_b} .$$

Let us turn to the study of the question of the full radiated energy and the efficiency of the transition electromagnetic pulse emitter based on a relativistic electron bunch. The full radiated energy is described by the expression

$$W = 2\pi\int_0^{\pi/2} \frac{W(\vartheta)}{do}\sin\vartheta d\vartheta . \quad (98)$$

Below we restrict ourselves to the limiting case of a short and thin beam (91), which can be analyzed analytically. In this limiting case, one can use expressions (93), (94) for the radiated energy per unit solid angle. As a result, for the total radiated energy, we obtain the following expression

$$W = \pi\sqrt{2\pi} W_0 \int_0^{\pi/2} \frac{\sin^3\vartheta}{\left(1-\beta_0^2\cos^2\vartheta\right)^2\sqrt{\tau_b^2+\rho_b^2\sin^2\vartheta}} d\vartheta .$$

This integral can be calculated. As a result, we obtain

$$W = \pi\sqrt{2\pi}\frac{W_0}{\rho_b} y^2 \frac{d}{dy}\left[\frac{y-1}{\sqrt{y(\alpha_b^2+1-y)}} Ln(y)\right], \quad (99)$$

$$Ln(y) = \ln\left[\frac{\alpha_b\sqrt{y}+\sqrt{\alpha_b^2+1-y}}{\sqrt{(\alpha_b^2+1)(y-1)}}\right].$$

Here $y = 1/\beta_0^2 > 1$.

These expressions are true at $\alpha_b > (\beta_0\gamma_0)^{-1}$. If a stronger condition $\alpha_b >> (\beta_0\gamma_0)^{-1}$ is satisfied, then formula (99) is simplified and takes the form

$$W = \sqrt{\frac{2}{\pi}}\frac{Q^2}{ct_b}\left[\ln\left(\frac{2\alpha_b\gamma_0}{\sqrt{\alpha_b^2+1}}\right) - \frac{1}{2}\right].$$

Within the considered approximation, the radiated energy of the transition pulse does not depend on the plasma density and grows weakly (logarithmically) with an increase of the relativistic factor of the bunch.

Let us define the efficiency of the emitter $\kappa_b$ based on transition radiation as the ratio of the radiated energy (98) to the kinetic energy of the bunch $W_{kin} = N_0 mc^2(\gamma_0 - 1)$, where $N_0 = Q/e$ is the number of particles in the bunch. Then for the efficiency we obtain the expression

$$\kappa_b = \frac{1}{\pi}\sqrt{\frac{2}{\pi}} N_0 \frac{r_{cl}}{ct_b}\frac{1}{\gamma_0 - 1}\ln\left(\frac{2\alpha_b\gamma_0}{\sqrt{\alpha_b^2+1}}\right),$$

where $r_{cl} = e^2/mc^2$ is the classical radius of the electron.

## CONCLUSION

In the present work, the process of excitation of a transition electromagnetic field by an electron bunch in a plasma half-space is investigated. The total electromagnetic field contains three components. First of all, in the volume of the plasma, the electron bunch will excite potential longitudinal plasma oscillations. The field of plasma oscillations includes a wake plasma wave, which has the same structure as in an unbounded plasma, as well as a clot of transition plasma oscillations localized in the vicinity of the plasma boundary. The amplitude of these oscillations on the plasma boundary is exactly equal to the amplitude of the wake plasma wave, but opposite in phase. With increase of distance from the region of injection of the electron bunch in plasma the amplitude of transition plasma oscillations continuously decreases and vanish. Note that transition plasma oscillations can play the role of a plasma electric undulator. When wake waves are excited in a plasma by multi-bunch systems, the effect of plasma FEL will take place [21-22], and high-frequency electromagnetic radiation will appear at a frequency $\omega = 2\gamma_0^2\omega_p$ .

The presence of a perfectly conducting plasma boundary also distorts the picture of the own quasi-static electromagnetic field of a relativistic electron bunch.

Induced on the perfectly conducting plane that bounded plasma, currents and charges excite the transition quasi-static electromagnetic field. This field is equivalent to the field of a positively charged mirror image of an electron bunch. The transition quasi-static field is also localized in the vicinity of the plasma boundary and has a pulsed character. The value of this field decreases as the electron bunch moves away from the boundary into the plasma volume.

And, finally, the full electromagnetic field contains a transition electromagnetic pulse excited by the electron bunch at the moment of its crossing the plasma boundary. The form of the transition pulse and its energy depend on the direction of radiation. An electromagnetic pulse has a sharp leading edge, its duration (longitudinal size), which is determined mainly by the duration of the electron bunch itself. The high-frequency component of the electromagnetic pulse is concentrated in the region of the leading edge. With distance from the leading edge to the injection plane of electron bunch, the frequency of the electromagnetic pulse decreases, and the wavelength, accordingly, increases. In the case of a relativistic electron bunch, the maximum of the radiation pattern is at small angles relatively to the direction of motion of the bunch.